\documentclass[conference]{IEEEtran}
\usepackage{graphicx}
\usepackage{tabularx}

\usepackage[T1]{fontenc}
\usepackage[utf8]{inputenc}
\usepackage[english]{babel}

\usepackage{csquotes}
\usepackage{nicefrac}
\usepackage{afterpage}
\usepackage{needspace}
\usepackage[textsize=scriptsize]{todonotes}
\usepackage{booktabs}
\usepackage{tikz}
\usepackage{quantikz}
\usetikzlibrary{arrows,positioning} 
\usetikzlibrary{shapes.geometric, positioning}
\usetikzlibrary{fit, quotes}
\usetikzlibrary{shapes.geometric, positioning}
\usetikzlibrary{automata,arrows}
\usepackage{flushend}
\usepackage{hyperref}
\usepackage{soul}
\usepackage{amssymb}
\usepackage{csvsimple}
\usepackage{amsthm}
\usepackage{lipsum}
\usepackage{varwidth}
\usepackage{utfsym}
\usepackage{fontawesome5}
\usepackage{multicol}
\usepackage{listings}
\usepackage{adjustbox}
\usepackage{float}
\usepackage{placeins}
\usepackage{yquant}
\usepackage{pgfplots}
\usepackage[detect-all=true]{siunitx}
\usepackage{etoolbox}
\usepackage{bbm}
\usepackage{bm}
\usepackage{balance}

\robustify\bfseries

\makeatletter
\let\MYcaption\@makecaption
\makeatother
\usepackage[font=footnotesize]{subcaption}
\makeatletter
\let\@makecaption\MYcaption
\makeatother

\newtheorem{example}{Example}

\usepackage[style=ieee, maxnames=9, minnames=6, date=year, doi=false, isbn=false, url=false, backend=biber, sortcites]{biblatex}
\AtEveryBibitem{
  \clearname{editor}
  \clearfield{series}
  \clearfield{isbn}
  \clearfield{issn}
  \clearfield{volume}
  \clearfield{number}
  \clearfield{pages}
  \clearfield{doi}
}

\newif\ifdoubleblind

\makeatletter
\newcommand{\redacted}[2][]{
  \ifdoubleblind
    \if\relax\detokenize{#1}\relax
      [redacted for double-blind review]%
    \else
      #1%
    \fi
  \else
    #2%
  \fi
}

\newcommand{\redactedHyper}[1]{
  \ifdoubleblind \else#1\fi
}

\newcommand{\nothing}{}
\makeatother

\definecolor{darkgreen}{RGB}{0, 150, 0}
\lstdefinelanguage{mqt-qasm}{
  keywords={include, qreg, creg, gate, measure, reset, barrier, if, opaque},
  keywordstyle=\color{blue}\bfseries,
  ndkeywords={x, h, cccx, ccx, cx, z, oracle, diffusion},
  ndkeywordstyle=\color{darkgray}\bfseries,
  identifierstyle=\color{black},
  sensitive=false,
  comment=[l]{//},
  morecomment=[s]{/*}{*/},
  commentstyle=\color{darkgreen}\ttfamily,
  stringstyle=\color{red}\ttfamily,
  morestring=[b]",
  morestring=[b]',
  basicstyle=\ttfamily\footnotesize,
  numbers=left,
  numberstyle=\tiny\color{gray},
  stepnumber=1,
  numbersep=5pt,
  showspaces=false,
  showstringspaces=false,
  showtabs=false,
  tabsize=2,
  breaklines=true,
  breakatwhitespace=false,
  captionpos=b,
  frame=single,
  alsoletter=-,
  lineskip=0pt,
}

\lstdefinestyle{qasm}{
    commentstyle=\color{darkgreen},
    keywordstyle=\color{blue},
    numberstyle=\tiny\color{gray},
    stringstyle=\color{red},
    basicstyle=\ttfamily\footnotesize,
    breakatwhitespace=false,         %
    breaklines=true,                 %
    captionpos=t,                    %
    keepspaces=true,                 %
    numbers=left,                    %
    numbersep=5pt,                   %
    showspaces=false,                %
    showstringspaces=false,          %
    showtabs=false,                  %
    tabsize=2,                       %
    frame=single,                    %
}
\newfloat{lstfloat}{htbp}{lop}
\floatname{lstfloat}{\footnotesize Listing}

\usepackage{hyperref}
\usepackage{makecell}

\doubleblindfalse
\renewcommand{\baselinestretch}{0.96}

\hypersetup{
	pdftitle={A Framework for the Efficient Evaluation of Runtime Assertions on Quantum Computers},                      %
	pdfsubject={},                    %
	pdfauthor={\redactedHyper{Damian Rovara, Lukas Burgholzer, Robert Wille}}  %
}

\newcommand{\chisquare}{power-divergence~}
\newcommand{\qn}[1]{q$_\texttt{#1}$}

\begin{document}

\renewcommand*{\figureautorefname}{Fig.}
\renewcommand*{\sectionautorefname}{Section}
\renewcommand*{\subsectionautorefname}{Section}
\def\exampleautorefname{Example}

\title{A Framework for the Efficient Evaluation of Runtime Assertions on Quantum Computers}

\author{
  \redacted[\nothing]{
\IEEEauthorblockN{Damian Rovara\IEEEauthorrefmark{1}\hspace*{1.5cm}Lukas Burgholzer\IEEEauthorrefmark{1}\IEEEauthorrefmark{2}\hspace*{1.5cm}Robert Wille\IEEEauthorrefmark{1}\IEEEauthorrefmark{3}\IEEEauthorrefmark{2}}
\IEEEauthorblockA{\IEEEauthorrefmark{1}Chair for Design Automation, Technical University of Munich, Germany}
\IEEEauthorblockA{\IEEEauthorrefmark{2}Munich Quantum Software Company GmbH, Garching near Munich, Germany}
\IEEEauthorblockA{\IEEEauthorrefmark{3}Software Competence Center Hagenberg GmbH (SCCH), Austria}
\IEEEauthorblockA{\href{mailto:damian.rovara@tum.de}{damian.rovara@tum.de}\hspace*{1.5cm}\href{mailto:lukas.burgholzer@tum.de}{lukas.burgholzer@tum.de}\hspace*{1.5cm}\href{mailto:robert.wille@tum.de}{robert.wille@tum.de}\\
\url{https://www.cda.cit.tum.de/research/quantum}}

  }
  }

\maketitle

\begin{abstract}
The continuous growth of quantum computing and the increasingly complex quantum programs resulting from it lead to unprecedented obstacles in ensuring program correctness. 
Runtime assertions are, therefore, becoming a crucial tool in the development of quantum programs.
They assist developers in the debugging process and help to test and verify the program.
However, while assertions can be implemented in a straightforward manner on classical computers, physical limitations of quantum computers pose considerable challenges for the evaluation of quantum assertions.
Access to the quantum state of a program is limited, execution time is expensive and noise can significantly distort measurement outcomes.
To address these problems, this work proposes a framework that assists developers in the evaluation of runtime assertions on real quantum computers.
It translates a variety of assertions into sets of measurements, reduces execution overhead where possible and evaluates the measurement results after the execution even in the presence of noise.
This approach substantially aids developers in the debugging process, enabling efficient assertion-driven debugging even in large programs.
The proposed framework is available as an open-source implementation at \redacted{\mbox{\url{https://github.com/munich-quantum-toolkit/debugger}}}.
\end{abstract}

\section{Introduction}
\label{sec:introduction}

Testing and debugging quantum programs is becoming an increasingly prominent area of research~\cite{dimatteo2024, metwalli2024, huang2019, miranskyy2021, garciadelabarrera2023}.
This includes the formal verification of programs~\cite{ying2012, liu2019}, equivalence checking~\cite{burgholzer2021}, but also methods adapted from classical debugging.
Runtime assertions, in particular, are a classical method experiencing a surge in popularity in quantum computing~\cite{huang2019a, li2020, li2014, liu2021, liu2020,rovara2024assertionrefinement}.
They allow developers to state expected properties of the program at any point during execution that are then verified dynamically.
This helps in narrowing down possible error locations and detecting changes that break previously functioning parts of the implementation.

However, while assertions have proven to be a powerful tool for classical debugging and testing~\cite{rosenblum1995}, similar methods in quantum programs are still severely limited.
Measuring quantum states in superpositions yields probabilistic results and the high frequency of gate and measurement errors on \mbox{state-of-the-art} quantum computers often falsify results.
While simulations can be used to circumvent these issues for smaller problems, the rapid growth of the field of quantum computing and its popular use cases requires a more scalable approach.
As quantum algorithms are starting to be applied to classical problems such as optimization problems, the same mechanisms that were previously used to guarantee the correctness of classical solutions are also desired for quantum approaches.

While initial methods for the evaluation of assertions on quantum computers have already been proposed, their usability in realistic situations is still sub-par.
Especially when executing quantum programs with a large number of assertions, as is often the case on the classical side, they quickly consume a large amount of expensive computational resources.
Furthermore, while classical assertions are often applied to verify all kinds of properties, quantum assertions typically only have a limited expressibility, allowing to test only for specific states.

To address these issues, this work proposes an automated framework for the evaluation of assertions on quantum computers.
It automatically translates abstract assertion specifications into measurements and splits the program into multiple slices to efficiently handle programs that employ multiple assertions. %
The generated slices can executed on a real quantum computer, producing a set of measurement outcomes that are used by the proposed framework to determine whether the assertions are satisfied.
By performing all of these steps automatically, the developer can work solely with high-level abstractions of the individual assertions rather than requiring to manually add the required measurements.
Moreover, all automated steps employ additional optimizations to minimize the overhead introduced for assertion evaluation.

Evaluations confirm that the proposed framework detects errors with high certainty in small problem instances on the example of the hardware-efficient ansatz, a popular building block for many applications of \emph{Noisy Intermediate Scale Quantum Computers}~(NISQ).
Additionally, the evaluations show that the employed methods are scalable even to larger instances where classical simulations become prohibitive.

The remainder of this work is structured as follows:
\autoref{sec:background} introduces quantum assertions and the specific assertion types considered in this work on an example program.
\autoref{sec:motivation} then illustrates the main challenges of evaluating assertions on real quantum computers and suggests an approach to address them.
\autoref{sec:proposed-framework} then introduces the proposed framework and illustrates its main components in more detail.
To evaluate the proposed framework, \autoref{sec:evaluation} then shows the results obtained from applying it to quantum programs of different sizes.
Finally, \autoref{sec:conclusion} concludes the work.

\section{Background}
\label{sec:background}

This section introduces key concepts related to the testing and debugging of quantum programs using assertions as well as further reviews related work in the field.

\subsection{Quantum Assertions}

Assertions are powerful methods for the runtime verification of programs. 
They are commonly employed in classical program verification~\cite{rosenblum1995, stallman2018}, assisting in debugging and regression testing~\cite{yoo2012, wong1997}.
An assertion enforces whether desired properties of the program state are satisfied during the execution of the program.
Many programming languages support the use of assertions as instructions that can be directly added to the program's source code.
In the development of quantum programs, they can similarly be employed to test for specific properties of the underlying quantum state.

In this work, we mainly focus on two categories of assertions: %
\begin{enumerate}
    \item \emph{Equality assertions} require the state of a given subset of qubits to be exactly equal to a given state representation. 
    In this context, the state may be provided in various ways, such as an explicit state representation or a circuit that prepares the desired state. 
    This type of assertions provides a rigid and precise control over the desired states that can often be employed at the end of a program to evaluate whether the final result is correct.
    
    \item \emph{Superposition assertions} require a given subset of qubits to be in a superposed state, i.e., these qubits must not be in a computational basis state.
    They are more flexible than equality assertions and can be effectively employed at points in the program where the exact state is not certain, such as for pre-condition checks for functions.
\end{enumerate}
This selection of assertions allows developers to specify the expected state precisely when high-detail information is available, as well as more leniently, when the program should instead be tested on a more general level.

\begin{example}
    \autoref{lst:bv} shows a simple implementation of the Bernstein-Vazirani algorithm~\cite{bernstein1997} using the OpenQASM~\cite{cross2022openqasm} standard that has been extended by the two assertion types introduced above.
    It will be used as a running example for the remainder of this work.
    Line~2 expresses that the qubit corresponding to the quantum variable \texttt{y} is expected to be in a superposition, i.e., its state is described as $\alpha_0\ket{0} + \alpha_1\ket{1}$ with $\alpha_0,\alpha_1 \neq 0$. 
    Line~13 and \mbox{Lines 14 to 18} specify two equality assertions where the former requires the qubit register \texttt{q} to be in the state $\ket{101}$ and the latter requires the qubit \texttt{anc} to be in the state $\ket{-}$, as expressed by the provided subcircuit.
\end{example}

\begin{lstfloat}[t]
\begin{lstlisting}[language=mqt-qasm,caption={An implementation of the Bernstein-Vazirani algorithm that employs assertions to verify the program's correctness.},label=lst:bv, escapeinside={(*}{*)}]
gate oracle x0, x1, x2, y {
    assert-sup y;
    cx x0, y;
    cx x2, y;
}

qreg q[3];
qreg anc[1];
x anc[0];
h q; h anc[0];
oracle q[0], q[1], q[2], anc[0];
h q;
assert-eq q = (*$|$*)101(*$\rangle$*);
assert-eq anc[0] {
    qreg t[1];
    h t[0];
    z t[0];
}
\end{lstlisting}
\end{lstfloat}

\subsection{Related Work}

In 2019, Huang et al.~\cite{huang2019a} proposed initial methods for the use of assertions to validate quantum programs.
These \emph{statistical assertions} allow developers to verify that the state of a quantum program is a speicifc computational basis state, in a superposition, or contains a set of entangled qubits.
This is achieved by running a $\chi^2$ test on measurement results obtained from running a program to verify whether they match the expected state.

However, as statistical assertions require possibly destructive measurements, alternatives were pursued~\cite{li2020, li2014, liu2021, liu2020} that leave the underlying quantum state intact.
In 2020, Li et al.~\cite{li2020} proposed \emph{Proq}, a runtime assertion scheme that uses \emph{projective measurements} that do not affect the underlying state.
To evaluate these assertions, additional instructions are added before measuring the state, transforming it to $\ket{\mathbf{0}}$ which can be measured without affecting the program.
While this method allows multiple assertions to be evaluated within one execution, it is limited to testing states for which this transformation can be derived efficiently.

Furthermore, both of these methods are severely impacted by the limitations of near-term quantum computers:
The additional instructions introduced by \emph{Proq} may greatly increase the depth of the executed circuit, leading to a decreasing fidelity on noisy devices.
Even for shallow circuits, invalid measurement results caused by the high level of noise can substantially skew the results of statistical evaluations.

To avoid the effects of noise, several approaches have been proposed to evaluate assertions on simulated quantum computers~\cite{rovara2024debugging,bauer-marquart2023}.
While these methods eliminate most of the limitations inherent to physical quantum computers, the computational complexity of classical quantum circuit simulation limits programs that may be verified using these methods in their size.

\section{Motivation and General Idea}
\label{sec:motivation}

This section highlights the challenges of assertion-based quantum debugging on real devices.
To circumvent these challenges, we then propose a general idea for the efficient use of assertions on real quantum computers.

\newcommand{\codescale}{0.6}
\newcommand{\codewidth}{0.30\textwidth}
\newcommand{\blockwidth}{0.05\textwidth}
\newcommand{\overlapSmall}{\hspace{-0.5cm}}

\begin{figure*}[h!]
    \centering
    \hspace{-3.75cm}
    \begin{subfigure}[t]{\codewidth}
        \centering
        \raisebox{3.75cm}{
            \scalebox{\codescale}[\codescale]{
                \begin{minipage}{\textwidth}
                    \input{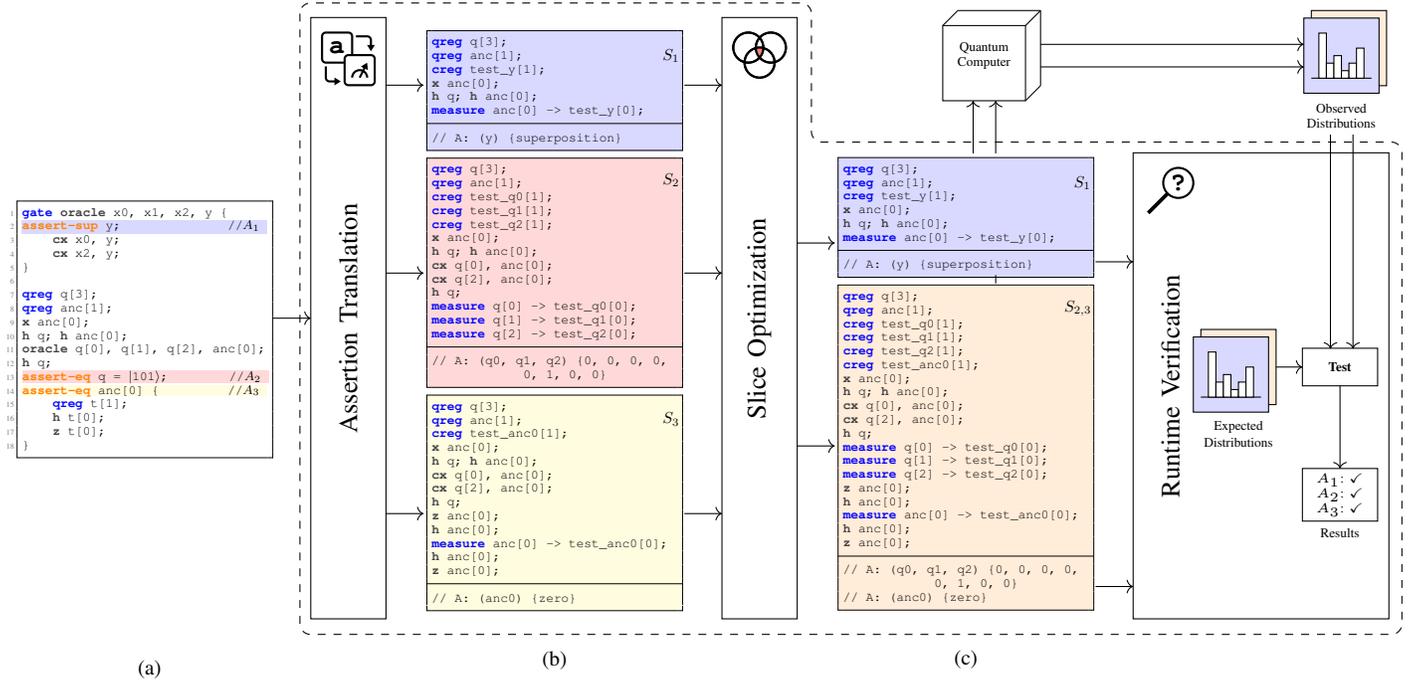}
                \end{minipage}
            }
        }
        \vspace{0.31cm}
        \caption{}
        \label{fig:outline-in}
    \end{subfigure}
    \hspace{-1.33cm}
    \begin{subfigure}[t]{\blockwidth}
        \centering
        \newcommand{\dyTranslation}{0.4}
\begin{tikzpicture}
    \draw[->] (-0.5, 4) -- (0, 4);
    \draw (0, 0) rectangle (1, 8);
    \node[rotate=90] at (0.5,4) {Assertion Translation};
    \draw[->] (1.0, 7.1) -- (1.5, 7.1);
    \draw[->] (1.0, 4.6) -- (1.5, 4.6);
    \draw[->] (1.0, 1.4) -- (1.5, 1.4);

    \draw[thick, rounded corners=0.5mm] (0.15, 7 + \dyTranslation) rectangle (0.55, 7.4 + \dyTranslation);
    \draw[thick, rounded corners=0.5mm,fill=white] (0.45, 7.1 + \dyTranslation) rectangle (0.85, 6.7 + \dyTranslation);
    \draw[thick, rounded corners=0.5mm,-{Triangle[scale=0.5]}] (0.2, 6.95 + \dyTranslation) -- (0.2, 6.75 + \dyTranslation) -- (0.4, 6.75 + \dyTranslation);
    \draw[thick, rounded corners=0.5mm,{Triangle[scale=0.5]}-] (0.8, 7.15 + \dyTranslation) -- (0.8, 7.35 + \dyTranslation) -- (0.6, 7.35 + \dyTranslation);
    \node at (0.35, 7.2 + \dyTranslation) {\textbf{\texttt{a}}};

    \draw[thick] (0.75, 6.8 + \dyTranslation) arc[start angle=0, end angle=180, radius=1mm];
    \draw[thick, -{Triangle[scale=0.5]}] (0.65, 6.8 + \dyTranslation) -- (0.75, 7.0 + \dyTranslation);
\end{tikzpicture}
    \end{subfigure}
    \begin{subfigure}[t]{\codewidth}
        \centering
        \raisebox{3.75cm}{
            \scalebox{\codescale}[\codescale]{
                \begin{minipage}{\textwidth}
                    \vspace{-0.1cm}
                    \begin{lstlisting}[language=mqt-qasm, escapeinside={(*}{*)}, numbers=none,backgroundcolor=\color{blue!15}]
qreg q[3];
qreg anc[1];
creg test_y[1];
x anc[0];
h q; h anc[0];
measure anc[0] -> test_y[0];

(*// A: (y) \{superposition\}*)
\end{lstlisting}
\begin{tikzpicture}[overlay]
    \node at (5.3, 2.6) {$S_1$};
    \def\ruleY{1.1}
    \draw (-0.12, \ruleY) -- (5.55, \ruleY);
\end{tikzpicture}
                    \vspace{-0.55cm}
                    \begin{lstlisting}[language=mqt-qasm, escapeinside={(*}{*)}, numbers=none,backgroundcolor=\color{red!15}]
qreg q[3];
qreg anc[1];
creg test_q0[1];
creg test_q1[1];
creg test_q2[1];
x anc[0];
h q; h anc[0];
cx q[0], anc[0];
cx q[2], anc[0];
h q;
measure q[0] -> test_q0[0];
measure q[1] -> test_q1[0];
measure q[2] -> test_q2[0];

(*// A: (q0, q1, q2) \{0, 0, 0, 0,*)
            0, 1, 0, 0}
\end{lstlisting}
\begin{tikzpicture}[overlay]
    \node at (5.3, 5.1) {$S_2$};
    \def\ruleY{1.4}
    \draw (-0.12, \ruleY) -- (5.55, \ruleY);
\end{tikzpicture}
                    \vspace{-0.55cm}
                    \begin{lstlisting}[language=mqt-qasm, escapeinside={(*}{*)}, numbers=none,backgroundcolor=\color{yellow!15}]
qreg q[3];
qreg anc[1];
creg test_anc0[1];
x anc[0];
h q; h anc[0];
cx q[0], anc[0];
cx q[2], anc[0];
h q;
z anc[0];
h anc[0];
measure anc[0] -> test_anc0[0];
h anc[0];
z anc[0];

(*// A: (anc0) \{zero\}*)
\end{lstlisting}
\begin{tikzpicture}[overlay]
    \node at (5.3, 4.75) {$S_3$};
    \def\ruleY{1.1}
    \draw (-0.12, \ruleY) -- (5.55, \ruleY);
\end{tikzpicture}
                \end{minipage}
            }
        }
        \caption{}
        \label{fig:outline-slices}
    \end{subfigure}
    \hspace{-1.25cm}
    \begin{subfigure}[t]{\blockwidth}
        \centering
        \newcommand{\dyOptimization}{0.4}
\usetikzlibrary{intersections}
\begin{tikzpicture}
    \draw (0, 0) rectangle (1, 8);
    \node[rotate=90] at (0.5,4) {Slice Optimization};
    \draw[->] (-0.5, 7.1) -- (0, 7.1);
    \draw[->] (-0.5, 4.6) -- (0, 4.6);
    \draw[->] (-0.5, 1.4) -- (0, 1.4);
    \draw[->] (1, 5.0) -- (1.5, 5.0);
    \draw[->] (1, 2.3) -- (1.5, 2.3);

    \coordinate (A) at (0.35, 7.2 + \dyOptimization);
    \coordinate (B) at (0.65, 7.2 + \dyOptimization);
    \coordinate (C) at (0.5, 7.0 + \dyOptimization);
    \def\radius{0.2}
    
    \draw[thick] (A) circle (\radius);
    \draw[thick] (B) circle (\radius);
    \draw[thick] (C) circle (\radius);
    \begin{scope}
        \clip (A) circle (\radius); %
        \clip (B) circle (\radius); %
        \clip (C) circle (\radius); %
        \fill[red!50] (A) circle (\radius); %
    \end{scope}
    \draw[thick] (A) circle (\radius);
    \draw[thick] (B) circle (\radius);
    \draw[thick] (C) circle (\radius);
\end{tikzpicture}
    \end{subfigure}
    \begin{subfigure}[t]{\codewidth}
        \centering
        \raisebox{3.75cm}{
            \scalebox{\codescale}[\codescale]{
                \begin{minipage}{\textwidth}
                    \vspace{2.7cm}

                    \begin{lstlisting}[language=mqt-qasm, escapeinside={(*}{*)}, numbers=none,backgroundcolor=\color{blue!15}]
qreg q[3];
qreg anc[1];
creg test_y[1];
x anc[0];
h q; h anc[0];
measure anc[0] -> test_y[0];

(*// A: (y) \{superposition\}*)
\end{lstlisting}
\begin{tikzpicture}[overlay]
    \node at (5.3, 2.6) {$S_1$};
    \def\ruleY{1.1}
    \draw (-0.12, \ruleY) -- (5.55, \ruleY);
\end{tikzpicture}
                    \vspace{-0.55cm}
                    \begin{lstlisting}[language=mqt-qasm, escapeinside={(*}{*)}, numbers=none,backgroundcolor=\color{orange!15}]
qreg q[3];
qreg anc[1];
creg test_q0[1];
creg test_q1[1];
creg test_q2[1];
creg test_anc0[1];
x anc[0];
h q; h anc[0];
cx q[0], anc[0];
cx q[2], anc[0];
h q;
measure q[0] -> test_q0[0];
measure q[1] -> test_q1[0];
measure q[2] -> test_q2[0];
z anc[0];
h anc[0];
measure anc[0] -> test_anc0[0];
h anc[0];
z anc[0];

(*// A: (q0, q1, q2) \{0, 0, 0, 0,*)
            0, 1, 0, 0}
(*// A: (anc0) \{zero\}*)
\end{lstlisting}
\begin{tikzpicture}[overlay]
    \node at (5.2, 7.2) {$S_{2,3}$};
    \def\ruleY{1.7}
    \draw (-0.12, \ruleY) -- (5.55, \ruleY);
\end{tikzpicture}
                \end{minipage}
            }
        }
        \caption{}
        \label{fig:outline-optimized}
    \end{subfigure}
    \hspace{-1.25cm}
    \begin{subfigure}[t]{\blockwidth}
        \centering
        \newcommand{\dyVerification}{-1.4}
\begin{tikzpicture}
    \draw (0, 0) rectangle (3.4, 6.2);
    \node[rotate=90] at (0.5,3.1) {Runtime Verification};
    \draw[->] (-0.5, 4.75) -- (0, 4.75);
    \draw[->] (-0.5, 0.43) -- (0, 0.43);
    \def\dxExp{-0.2}
    \def\dyExp{0.35}

    \draw[fill=orange!15] (1.1 + \dxExp, 2.5 + \dyExp) rectangle(2.1 + \dxExp, 3.5 + \dyExp);
    \draw[fill=blue!15] (1 + \dxExp, 2.4 + \dyExp) rectangle(2 + \dxExp, 3.4 + \dyExp);
    \draw (1.1 + \dxExp, 2.6 + \dyExp) -- (1.9 + \dxExp, 2.6 + \dyExp);
    \draw[fill=white] (1.2 + \dxExp, 2.6 + \dyExp) rectangle (1.3 + \dxExp, 3.2 + \dyExp);
    \draw[fill=white] (1.3 + \dxExp, 2.6 + \dyExp) rectangle (1.4 + \dxExp, 2.8 + \dyExp);
    \draw[fill=white] (1.4 + \dxExp, 2.6 + \dyExp) rectangle (1.5 + \dxExp, 2.9 + \dyExp);
    \draw[fill=white] (1.5 + \dxExp, 2.6 + \dyExp) rectangle (1.6 + \dxExp, 2.7 + \dyExp);
    \draw[fill=white] (1.6 + \dxExp, 2.6 + \dyExp) rectangle (1.7 + \dxExp, 2.8 + \dyExp);
    \draw[fill=white] (1.7 + \dxExp, 2.6 + \dyExp) rectangle (1.8 + \dxExp, 3.0 + \dyExp);
    \node at (1.6 + \dxExp, 2.2 + \dyExp) {\tiny Expected};
    \node at (1.6 + \dxExp, 2.0 + \dyExp) {\tiny Distributions};

    \draw[->] (2.1 + \dxExp, 3.0 + \dyExp) -- (2.45 + \dxExp, 3.0 + \dyExp);

    \def\dxTest{1.15}
    \def\dyTest{-1.4}
    \draw (1.1 + \dxTest, 4.5 + \dyTest) rectangle (2.1 + \dxTest, 5.0 + \dyTest);
    \node at (1.6 + \dxTest, 4.75 + \dyTest) {\tiny \textbf{Test}};
    \draw (1.1 + \dxTest, 3.4 + \dyTest) rectangle (2.1 + \dxTest, 2.7 + \dyTest);
    \node at (1.6 + \dxTest, 2.75 + \dyTest + 0.5) {\tiny $A_1$: \checkmark};
    \node at (1.6 + \dxTest, 2.55 + \dyTest + 0.5) {\tiny $A_2$: \checkmark};
    \node at (1.6 + \dxTest, 2.35 + \dyTest + 0.5) {\tiny $A_3$: \checkmark};
    \node at (1.6 + \dxTest, 2.05 + \dyTest + 0.5) {\tiny Results};
    \draw[->] (1.6 + \dxTest, 4.5 + \dyTest) -- (1.6 + \dxTest, 3.4 + \dyTest);

    \draw[thick] (0.6, 7.2 + \dyVerification) circle (0.2);
    \draw[line width=0.5mm] (0.2, 6.8 + \dyVerification) -- (0.46, 7.06 + \dyVerification);
    \node at (0.6, 7.2 + \dyVerification) {\textbf{\texttt{?}}};

\end{tikzpicture}
    \end{subfigure}
    \begin{tikzpicture}[overlay]
    \draw[->] (2.20 - 0.1, 6.45) -- (2.20 - 0.1, 3.6);
    \draw[->] (2.50 - 0.1, 6.45) -- (2.50 - 0.1, 3.6);
    \draw[->] (-2.65, 6.2) -- (-2.65, 6.9);
    \draw[->] (-2.35, 6.2) -- (-2.35, 6.9);
    \draw (-2.35, 4.47) -- (-2.35, 4.58);
    \draw[dashed, rounded corners] (-11.6, 8.2) -- (-11.6, -0.2) -- (3.05, -0.2) -- (3.05, 6.35) -- (-4.8, 6.35) -- (-4.8, 8.2) -- cycle;
\end{tikzpicture}
    \begin{tikzpicture}[overlay]
    \def\dy{0.4}
    \node at (-2.75, 7.2 + \dy) {\tiny Quantum};
    \node at (-2.75, 7.0 + \dy) {\tiny Computer};
    \draw (-3.3, 7.5 + \dy) rectangle (-2.2, 6.5 + \dy);
    \draw (-3.3, 7.5 + \dy) -- (-3.1, 7.7 + \dy);
    \draw (-2.2, 7.5 + \dy) -- (-2.0, 7.7 + \dy);
    \draw (-2.2, 6.5 + \dy) -- (-2.0, 6.7 + \dy);
    \draw (-3.1, 7.7 + \dy) -- (-2.0, 7.7 + \dy);
    \draw (-2.0, 6.7 + \dy) -- (-2.0, 7.7 + \dy);

    \def\dxObs{2.0}

    \draw[->] (-2.0, 7.25 + \dy) -- (-0.5 + \dxObs, 7.25 + \dy);
    \draw[->] (-2.0, 6.95 + \dy) -- (-0.5 + \dxObs, 6.95 + \dy);
    \draw[fill=orange!15] (-0.4 + \dxObs, 6.7 + \dy) rectangle (0.6 + \dxObs, 7.7 + \dy);
    \draw[fill=blue!15] (-0.5 + \dxObs, 6.6 + \dy) rectangle (0.5 + \dxObs, 7.6 + \dy);
    \draw (-0.4 + \dxObs, 6.8 + \dy) -- (0.4 + \dxObs, 6.8 + \dy);
    \draw[fill=white] (-0.3 + \dxObs, 6.8 + \dy) rectangle (-0.2 + \dxObs, 7.4 + \dy);
    \draw[fill=white] (-0.2 + \dxObs, 6.8 + \dy) rectangle (-0.1 + \dxObs, 7.0 + \dy);
    \draw[fill=white] (-0.1 + \dxObs, 6.8 + \dy) rectangle (-0.0 + \dxObs, 7.1 + \dy);
    \draw[fill=white] (-0.0 + \dxObs, 6.8 + \dy) rectangle (0.1 + \dxObs, 6.9 + \dy);
    \draw[fill=white] (0.1 + \dxObs, 6.8 + \dy) rectangle (0.2 + \dxObs, 7.0 + \dy);
    \draw[fill=white] (0.2 + \dxObs, 6.8 + \dy) rectangle (0.3 + \dxObs, 7.2 + \dy);
    \node at (0 + \dxObs, 6.4 + \dy) {\tiny Observed};
    \node at (0 + \dxObs, 6.2 + \dy) {\tiny Distributions};

\end{tikzpicture}
    \caption{The workflow of the proposed framework, split into its three core steps.}
    \label{fig:outline}
    \vspace{-0.6cm}
\end{figure*}

\subsection{Considered Problem}
\label{sec:considered-problem}

Tools and frameworks that aid in the development of quantum programs are becoming increasingly popular.
Efficient simulators allow developers to test small instances of their programs on classical hardware before running them on real quantum devices.
These simulators typically also provide additional features such as debugging tools and state access during the execution of the program.
Such features also allow assertions to be tested during the simulation of the program without considerable overhead~\cite{rovara2024debugging, metwalli2024}.

However, programs that can be tested on simulated quantum computers are strictly limited in their size.
Futhermore, the benefits of assertion testing on simulated devices do not directly translate to real quantum hardware that comes with further physical constraints.
These constraints introduce limitations that make debugging and testing quantum programs particularly challenging.

More precisely, the evaluation of assertions on real quantum computers suffers from three core issues:
\begin{enumerate}
\item \emph{Limited state access:} Real quantum devices do not provide the same insight in the current state of the program that many simulators can.
Instead, the state of the program can only be estimated through measurements.
However, the exact extraction of a quantum state through measurements is known to be a hard problem.
Furthermore, measurements may cause the measured state to collapse, affecting the correctness of the program in a way that is not a concern in classical testing, limiting the access to the underlying state even further.
\item \emph{Efficiency:} As obtaining the state of the program may require destructive measurements, the evaluation of multiple assertions often requires repeated executions of a program.
Due to the limited availability of quantum devices, the excessive use of these resources can be time-consuming and costly.
It is, therefore, important that special care is taken to keep the overhead introduced by evaluating assertions to a minimum.
\item \emph{Noise:} Finally, near-term quantum computers are still severely limited in their fidelity.
Especially the execution of large programs often carries a considerable degree of noise that may distort measurement results.
Therefore, fully relying on measurement outcomes to evaluate the correctness of assertions quickly leads to false positives.
Instead, an effective framework for quantum assertions should be as robust as possible to noise and only detect errors that are caused by anomalies in the program. 
\end{enumerate}

\subsection{General Idea}
\label{sec:general-idea}

To allow developers to efficiently test quantum programs on real devices, this work proposes solutions to directly address all of the above issues.
These solutions are then combined into a comprehensive framework that is illustrated in \autoref{fig:outline}.

It starts with a program specification that uses assertion to test for its correctness.
To evaluate an assertion on a real quantum computer, measurements must be performed to access the underlying state.
However, as such measurements collapse the state, execution cannot continue afterwards.
Therefore, the program is divided into multiple slices.
Each slice concerns only one assertion and includes only the measurements required for it.
Depending on the type of a considered assertion, additional operations may have to be introduced to allow the required degree of state access.
The proposed framework automatically handles the efficient addition of these operations to individual slices when necessary.
We refer to the process of translating a program containing assertions to a set of slices with measurements as \emph{assertion translation}.

\vspace{-0.1cm}

\begin{example}
Consider the Bernstein-Vazirani quantum program passed as input to the proposed framework in \autoref{fig:outline-in}. 
For each of the highlighted assertions, denoted as $A_1$, $A_2$, and $A_3$, a set of measurement instructions used to evaluate their correctness is generated.
The program is then split into three slices, $S_1$, $S_2$, and $S_3$ as illustrated in \autoref{fig:outline-slices}.
Each of the slices contains measurements of all target variables of the corresponding assertions on \mbox{Lines 2, 13, and 14} in \autoref{lst:bv}.
For the assertion on Line~14, additional operations are added to the slice $S_3$.
This process will be further explained later in \autoref{sec:assertion-translation}.
\end{example}

As each of the slices generated during assertion translation must be executed on physical quantum computers, a high number of slices causes severe execution overhead.
Using \emph{slice optimization}, we attempt to reduce the total number of slices.
This is achieved by grouping multiple assertions with measurements that do not interfere with each other into larger slices.
Additionally, assertions that are guaranteed to be satisfied based on previous assertions can be cancelled, reducing the number of slices even further.

\begin{example}
Consider the assertions $A_2$ and $A_3$ on \mbox{Lines 13 and 14} in \autoref{lst:bv}, corresponding to the slices $S_2$ and $S_3$ in \autoref{fig:outline-slices}.
As these assertions operate on disjoint sets of qubits and no target qubit is reused after the first assertion, these assertions do not interfere with each other.
Therefore, they can be concatenated into a single slice, $S_{2,3}$, as shown in \autoref{fig:outline-optimized}.
Conversely, while the assertions $A_1$ and $A_2$ on \mbox{Lines 2 and 13} also operate on disjoint qubit sets, several instructions targeting the qubit \texttt{y} still appear between the two assertions.
Therefore, it is not possible to guarantee that the measurement introduced to translate the assertion on Line~2 does not invalidate further operations and the slices cannot be concatenated.
\end{example}

The optimized slices are then executed on a quantum computer.
However, to accurately evaluate the assertions, multiple executions per slice are required.
These measurement outcomes are then utilized to perform \emph{runtime verification}, evaluating whether each assertion is satisfied.
To this end, we employ statistical tests to compare the observed measurement outcomes with the expected outcomes based on a given assertion.
Additionally, the expected fidelity of each slice is taken into account to mitigate possible influences of noise when evaluating an assertion.
Based on a desired confidence level provided by the user, the correctness of each assertion can be evaluated from the results of the employed tests.

\begin{example}
To statistically evaluate the correctness of the program, the slices $S_1$ and $S_{2,3}$ are executed on a quantum computer over multiple iterations.
The resulting measurement outcomes are then passed to the proposed framework.
Based on the provided device information, the proposed framework computes the expected fidelity of each slice and uses it to construct an expected distribution.
For the equality assertion on Line~13, for instance, a \chisquare\cite{cressie1984} test is then employed to determine whether the observed outcomes are consistent with the expected distribution.
If the test yields a $p$-value exceeding the user-defined threshold, the assertion is accepted.
\end{example}

Attempting to resolve all three of the above-mentioned issues at once inevitably involves a trade-off:
Established methods for detailed state access, such as state tomography algorithms, scale poorly in the number of qubits~\cite{vogel1989, cramer2010}.
Furthermore, high-confidence measurements in the presence of noise require an exceedingly large number of executions.
Crucially, this work proposes a comprehensive framework that aims to address this trade-off efficiently:
While its supported assertion types may not be able to evaluate every state exactly, they greatly limit the execution overhead for state access.
Similarly, by allowing the developer to limit the desired level of confidence the number of required executions is further reduced.
This allows the proposed framework to handle the evaluation of assertions on physical quantum computers in an efficient manner.

\section{Proposed Framework}
\label{sec:proposed-framework}

In this section, we describe the framework introduced above in further detail. 
In particular, we focus on the three core steps proposed in \autoref{sec:general-idea} required to evaluate quantum assertions at runtime on phyiscal devices:
\emph{assertion translation}, \emph{slice optimization}, and \emph{runtime verification}.
The framework, including all individual steps, is available as an open-source implementation at \redacted{\mbox{\url{https://github.com/munich-quantum-toolkit/debugger}}}.

\subsection{Assertion Translation}
\label{sec:assertion-translation}

To evaluate assertions on phyiscal quantum computers, they must first be translated into instructions that are understood by the device.
Generally, insight into the current state of a program during execution can only be achieved by measuring qubits.
Therefore, evaluating assertions requires the insertion of measurements into the program.
However, these measurements can collapse the program's state, forcing the execution to terminate.
To evaluate multiple assertions within a single quantum program, the program is, therefore, split into multiple \emph{slices}.
The $n$-th slice starts at the beginning of the program and ends at the $n$-th assertion.
Slices are then executed independently, allowing each assertion to be verified individually.

The following describes how each of the assertion types reviewed in \autoref{sec:background} can be translated to measurements.

\paragraph*{Superposition Assertions} 
A superposition assertion can be translated into a set of measurements where each qubit included in the assertion's target list is measured individually.  
In a noise-free environment, these measurements will provide multiple distinct results if and only if the original set of qubits was in a superposition state.

\begin{example}
\autoref{fig:outline-in} illustrates an input to the assertion translation procedure containing three assertions.
For this input, the first slice reaches from the beginning of the program to the superposition assertion on Line~2 after the call to the custom gate \texttt{oracle} is inlined.
For the creation of this slice, a classical variable \texttt{test\_y} is generated and the measurement outcome of~\texttt{anc[0]} is stored to it.
Slice $S_1$ in \autoref{fig:outline-slices}.
\end{example}

\paragraph*{Equality Assertions with Explicit State Specification}
Based on the expected state $\ket{\psi}$, we compute the expected probability $\left | \left \langle i | \psi \right \rangle \right |^2$ for each measurement outcome $\ket{i}$.
The observed probability distribution can then be compared with the expected distribution.
In particular, if the observed and expected distributions differ, then the current state of the system must differ from the state required by the assertion.
However, as the same probability distributions may result from different state amplitudes, this form of evaluation cannot definitively prove that the assertion is satisfied.
Instead, this method can only provide probabilistic evidence for the assertion.
While state tomography could instead offer a more precise comparison, it incurs higher measurement costs while still being unable to guarantee accuracy in some cases. 
The proposed approach favors efficiency with statistically robust evidence.

\begin{example}
Line~13 in \autoref{fig:outline-in} expresses an equality assertion on the qubit register \texttt{q}.
As this assertion targets three qubits, three new classical variables and three measurements are introduced to the corresponding program slice $S_2$.
The previous superposition assertion on Line~2 is not included in this slice.
\end{example}

\paragraph*{Equality Assertions with Circuit Specification}
Viewing the unitary operations applied in the given circuit specification as individual gates $g_i$ with $G = g_0\dots g_{|G| - 1}$, we can construct the inverse operation $G^\dagger = g^{-1}_{|G| - 1}\dots g^{-1}_0$.
If the assertion is satisfied, the current state of the system is then equal to $G \ket{\mathbf{0}}$.
The state of a system satisfying the assertion can, therefore, be reverted to the $\ket{\mathbf{0}}$ state by applying its inverse $G^\dagger$ at the place of the assertion.
In this case, we expect all qubits in the sampled distribution to be equal to $\ket{0}$.
This simplifies the verification process to just checking the $\ket{\mathbf{0}}$ state rather than a more complex arbitrary state.
After the measurement, $G$ can be applied once again to return the system to the correct state and the execution can be continued to the following assertion without requiring an additional slice to be created for it.
However, as the application of $G$ may lead to a large number of instructions in the resulting slice that will have to be compiled for the target architecture, the developer is also given the choice to not re-apply $G$ and instead stop the execution at this point.
This allows them to freely balance compilation and execution overhead with the number of slices.
This approach is in line with \emph{projection-based assertions} proposed by Li et al.~\cite{li2020}.

\begin{example}
    Line~14 in \autoref{fig:outline-in} expresses an equality assertion that provides a circuit representation for the desired state of the qubit \texttt{anc}.
    To translate this assertion, the provided circuit is first inverted, resulting in the operations \texttt{z} and \texttt{h}.
    These inverted operations are then added to the end of the slice, followed by a measurement of the \texttt{anc} qubit.
    The measurement outcome should be \texttt{\ket{0}} at this point of the program.
    Finally, to return the program to the correct state, the gates given by the assertion's provided circuit are applied once again.
    The resulting slice for this assertion is illustrated as $S_3$ in \autoref{fig:outline-slices}. 
\end{example}

Clearly, these methods are not equally applicable in all situations.
When testing equality assertions with explicit state specification, for instance, the chance of false negatives limits the accuracy of the assertions.
To use equality assertions based on circuit specification, on the other hand, the circuit to construct the expected state must be known.
However, this mixed approach allows both of these methods to be used interchangably according to the preference and available information of the developer.
Furthermore, for small instances, it is possible to translate between equality assertions with explicit state specifications and circuit specifications automatically.
This is because simulation methods can be applied to compute the expected state vector of a circuit and, conversely, solving the state preparation problem for a given explicit state allows the synthesis of the equivalent circuit.
However, as both of these methods scale poorly as the number of qubits increases, this translation can only be applied in simple cases.
\bigskip

The above methods have shown how assertions can be translated into program slices with measurements to be executed on quantum computers.
However, due to the high degree of repetition in the generated program slices, the creation of a large number of slices may lead to a disproportionate amount of overhead.
If every assertion requires re-executing the entire program up to that point, the number of actually executed instructions scales quadratically with the total number of instructions in the program.
Therefore, an approach for reducing the number of program slices that need to be executed is proposed next.

\subsection{Slice Optimization}

\begin{table}
    \caption{The conditions required for an assertion $A_1$ (rows) to imply a subsequent assertion $A_2$ (columns)}
    \label{tab:implications}
    \begin{tabular}{r|c|c}
        & Superposition & Equality \\ \hline
        Superposition & $Q_1 \subseteq Q_2$ & $\times$ \\
        Equality & \makecell{$Q = Q_1 \cap Q_2 \neq \emptyset$ and \\ $Q$ is in a superposition in $\ket{\psi_1}$} & \makecell{$Q_2 \subseteq Q_1$ and\\$\ket{\psi_1} = \ket{\psi_2} \otimes \ket{\phi}$ \\for some $\ket{\phi}$} \\
    \end{tabular}
\end{table}

To reduce the execution overhead when evaluating multiple assertions, we propose two optimization methods:

\paragraph*{Subset Canceling} For each assertion $A$, we denote $S(A)$ as the set of all quantum states that satisfy $A$.
Given two directly consecutive assertions $A_i$ and $A_j$, we say that $A_i$ implies $A_j$ if~$S(A_i) \subseteq S(A_j)$.
This is because, for any state $\ket{\psi} \in S(A_i)$, $\ket{\psi}$ satisfies $A_i$.
However, as $S(A_i) \subseteq S(A_j)$, $\ket{\psi}$ is also an element of $S(A_j)$ and, therefore, satisfies $A_j$.
Then, if~$A_i$~implies~$A_j$, the slice for $A_j$ can be eliminated safely.
This approach can only be applied, if the two considered assertions are directly consecutive, i.e., no further instruction appear between them.
Otherwise, the proposed framework attempts to move the later assertion closer to the first.
This is done by iteratively moving it over the previous instruction as long as the assertion commutes with it~\cite{rovara2024assertionrefinement}.
If this process can be repeated until the two assertions are directly consecutive, subset canceling can once again be applied.

The rules for subset canceling depend on the considered assertions.
A superposition assertion is defined by its target qubits $Q$, whereas an equality assertion is defined by both its target qubits $Q$ and its expected state $\ket{\psi}$.
Based on these defining factors, \autoref{tab:implications} lists all possible combinations of assertions that may be eliminated through subset canceling.

\begin{figure}
    \centering
    \begin{subfigure}{0.45\columnwidth}
        \centering
        \begin{tikzpicture}
            \draw (0, 0) circle [radius=1.2cm];
            \draw (1.8, 0) circle [radius=1.2cm];
            \draw (0.35, -0.4) circle [radius=0.65cm];
            \node at (0, 0.7) {\tiny $A_2$: \texttt{sup \qn{0},\qn{1}}};
            \node at (1.85, 0.7) {\tiny $A_3$: \texttt{sup \qn{1},\qn{2}}};
            \node[fill=white, inner sep =1pt] at (0.35, -0.3) {\tiny $A_1$: \texttt{sup \qn{0}}};
        \end{tikzpicture}
        \caption{\vspace{0.2cm}$S(A_1) \subseteq S(A_2) \not \subseteq S(A_3)$}
        \label{fig:venn-sup-sup}
    \end{subfigure}
    \begin{subfigure}{0.45\columnwidth}
        \centering
        \begin{tikzpicture}
            \draw (0, 0) ellipse [x radius=1.2cm, y radius=1.2cm];
            \draw (1.2, 0) ellipse [x radius=1.2cm, y radius=1.2cm];
            \node at (-0.5, 0.1) {\tiny $A_2$:};
            \node at (-0.5, -0.1) {\tiny \texttt{eq \qn{0}=$\ket{0}$}};
            \node at (1.8, 0.1) {\tiny $A_3$:};
            \node at (1.8, -0.1) {\tiny \texttt{eq \qn{1}=$\ket{1}$}};
            \node at (0.6, 0.1) {\tiny $A_1$: \texttt{eq}};
            \node at (0.6, -0.1) {\tiny \texttt{\qn{0},\qn{1}=$\ket{10}$}};
        \end{tikzpicture}

        \caption{\vspace{0.2cm}$S(A_1) = S(A_2) \cap S(A_3)$}
        \label{fig:venn-eq-eq}
    \end{subfigure}

    \begin{subfigure}{0.45\columnwidth}
        \centering
        \begin{tikzpicture}
            \draw (0, 0) circle [radius=1.2cm];
            \draw (0, -0.4) circle [radius=0.7cm];
            \node at (0, 0.7) {\tiny $A_2$: \texttt{sup \qn{0}}};
            \node at (0, -0.2) {\tiny $A_1$:};
            \node at (0, -0.4) {\tiny \texttt{eq \qn{0}=}$\ket{+}$};
        \end{tikzpicture}
        \caption{$S(A_1) \subseteq S(A_2)$}
        \label{fig:venn-eq-sup}
    \end{subfigure}
    \begin{subfigure}{0.45\columnwidth}
        \centering
        \begin{tikzpicture}
            \draw (0, 1) ellipse [x radius=1.2cm, y radius=0.7cm];
            \draw (0, 0) ellipse [x radius=1.2cm, y radius=0.7cm];
            \node at (0, 1.3) {\tiny $A_1$: \texttt{eq \qn{0} = }$\ket{\texttt{1}}$};
            \node at (0, -0.3) {\tiny $A_2$: \texttt{sup \qn{0},\qn{1}}};
        \end{tikzpicture}
        \caption{$S(A_1) \not \subseteq S(A_2)$}
        \label{fig:venn-sup-eq}
    \end{subfigure}
    \caption{The subset canceling interactions for different types of assertions.}
    \label{fig:venn}
\end{figure}
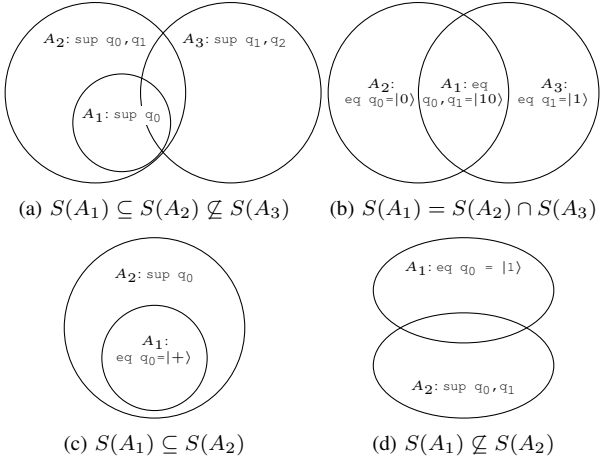

\begin{example}
\autoref{fig:venn} illustrates the interactions between different assertions and the sets of qubits that satisy them.
\autoref{fig:venn-sup-sup} and \autoref{fig:venn-eq-sup} show two cases of a larger, more general superposition assertion being implied by stricter assertions, according to the rules in \autoref{tab:implications}.
In both of these cases, the stricter assertion defines a subset of the outer assertion.
In particular, the stricter superposition assertion over $q_0$ implies that the qubit set $q_0, q_1$ is also in a superposition.
The equality assertion expecting $q_0 = \ket{+}$ also implies that $q_0$ is in a superposition.
\autoref{fig:venn-eq-eq} shows the interaction of multiple equality assertions. 
In particular, it shows how the stricter equality assertion over the qubits $q_0$ and $q_1$ implies both more general assertions over the individual qubits.
This is because $\ket{q_0q_1} = \ket{01}$ can only hold if $q_0 = \ket{0}$ and $q_1 = \ket{1}$.
Finally, \autoref{fig:venn-sup-eq} shows a case in which no subset canceling can be performed because the shared qubit $q_0$ is expected to be in the computational basis state $\ket{1}$ by the equality assertion.
\end{example}

Subset canceling can drastically reduce the overhead required for programs that contain a large number of repetitive assertions.
However, the total number of slices can be reduced even further by concatenating independent slices.

\paragraph*{Slice Concatenation} An assertion targeting a set of qubits $Q$ requires each qubit in $Q$ to be measured.
As such measurements may collapse the underlying state, the correctness of any further execution may be impeded.
However, if the measurement does not interfere with any remaining operations up to the following assertion, the slices required for the two assertions can be concatenated into a single slice.
This is the case if no qubit measured for the first assertion is used as the target of any other instruction afterwards.
Due to the \emph{deferred measurement principle}~\cite{nielsen2010}, the measurements will not impact the function of the following instructions and can remain in place.

Furthermore, measurements that do not lead to the collapse of the underlying state will never impact subsequent instructions and, therefore, always support concatenation.
This is the case if an equality assertion specifies a single computational basis state as the expected state.
As this implies the absence of superposition in the measured qubits, the measurement will not impact a correct state and execution can continue immediately.
Furthermore, all equality assertions with circuit specifications allow for projective measurements by uncomputing the state of the target qubits to $\ket{0}$.
As this state is not in a superposition, these measurements, similarly, do not cause a collapse of the qubits' states and, therefore, always allow slice concatenation.

\begin{figure}
    \centering
    \begin{subfigure}{0.55\columnwidth}
        \begin{lstlisting}[language=mqt-qasm, numbers=none]
h q[0];

h q[1];

assert-sup q[0], q[1];

cx q[1], q[2];

assert-sup q[2];
 
\end{lstlisting}
\begin{tikzpicture}[overlay]
    \draw[rounded corners=0.2cm,fill=orange!20] (1.75, 2.32) rectangle (3.85, 1.78);
    \draw[rounded corners=0.2cm,fill=blue!20] (1.5, 1.18) -- (2.4, 1.18) -- (2.4, 1.72) -- (1.5, 1.72);
    \draw[rounded corners=0.2cm,dashed,fill=blue!20] (1.5, 1.18) -- (0.43, 1.18) -- (0.43, 1.72) -- (1.5, 1.72);
    \node at (2.8, 2.05) {\footnotesize \texttt{q[0], q[1];}};
    \node at (1.45, 1.45) {\footnotesize \texttt{q[1], q[2];}};
\end{tikzpicture}
    \end{subfigure}
    \hfill
    \begin{subfigure}{0.4\columnwidth}
        \raisebox{0.5cm} {
        \begin{tikzpicture}
            \draw (0, 0) circle [radius=1.5cm];
            \draw[fill=blue!20] (0.3, -0.2) ellipse [x radius=1.1cm, y radius=0.6cm, rotate=90];
            \draw[fill=orange!20] (-0.1, 0.3) ellipse [x radius=1.2cm, y radius=0.6cm, rotate=30];
            \draw[dashed] (0.3, -0.2) ellipse [x radius=1.1cm, y radius=0.6cm, rotate=90];
            \node at (0.325, 0.3) {\footnotesize \texttt{q[1]}};
            \node at (-0.7, -0.1) {\footnotesize \texttt{q[0]}};
            \node at (0.325, -0.85) {\footnotesize \texttt{q[2]}};

            \node[rotate=35] at (-0.6, 1.1) {\footnotesize $S_{1,2}$};
            \node[rotate=0] at (-0.5, 0.45) {\footnotesize $S_1$};
            \node[rotate=0] at (0.65, -0.25) {\footnotesize $S_2$};
        \end{tikzpicture}
        }
    \end{subfigure}
\vspace{-0.5cm}

    \caption{As no instruction targets $q_1$ or $q_2$, the two slices $S_1$ and $S_2$ can be concatenated into one slice $S_{1, 2}$.}
    \label{fig:concat}
\end{figure}
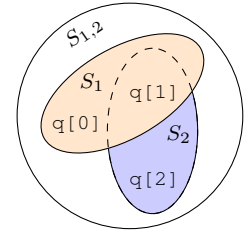

\begin{example}
Consider the program illustrated in \autoref{fig:concat}.  
Between the first and second assertion, a single instruction, \texttt{cx q1, q2} is located.
After translating the first assertion, the qubits \texttt{q0} and \texttt{q1} will be measured.
However, as \texttt{q1} is only used as a control qubit for the subsequnt \texttt{cx} instruction and \texttt{q0} is not reused at all, this measurement does not impact the results of the measurement of \texttt{q2} required to translate the second assertion.
Therefore, all of these measurements can be co-located in a single slice.
\end{example}

\subsection{Runtime Verification}

After assertion translation and slice optimization, a set of program slices is generated that can be executed on a quantum computer.
During execution, all qubits required for the verification of assertions are measured and the measurement results are collected.
This process is repeated over several iterations to produce a sampled distribution over all measured qubits.
By returning these measurement results to the proposed framework's \emph{runtime verification} scheme, the satisfaction of the assertions is verified automatically.

To this end, during the assertion translation step, each translated assertion is prepended with a metadata string specifying its evaluation criteria.
For \emph{superposition assertions} and \emph{equality assertions with circuit specification}, the preamble lists the names of all measurement results and indicates whether a superposition or an all-zero state is expected.
For \emph{equality assertions with explicit state specification}, the preamble instead indicates the expected probability distribution.

For superposition assertions, any state with a probability equal to $1$ indicates that the program state is not in a superposition and the assertion is rejected.
Similarly, for equality assertions with circuit specification, the assertion is rejeted if the observed distribution is different from $\ket{\mathbf{0}}$.
For equality assertions with explicit state specification, on the other hand, the porposed framework utilizes a \emph{\chisquare}~\cite{cressie1984} goodness-of-fit test to compare the sampled distribution with the expected distribution.
It allows the developer to propose any desired $p$-value that is used to determine whether an assertion is satisfied.
If the \chisquare test results in a $p$-value less than or equal to the desired value, the assertion is rejected.

However, as near-term quantum computers still suffer from a high level of noise that may cause measurements to result in unexpected values, this method is not always applicable.
To mitigate this issue, the proposed framework also allows for the specification of gate and measurement fidelities of the physical quantum computer.
Based on these fidelities, the framework computes the expected fidelity of each slice before evaluating the assertions.
This is achieved by multiplying the expected error rates of each gate applied in the slice, representing an estimate for the likelihood of the execution leading to the correct results.
It then modifies the expected distribution based on the slice fidelity by mixing it with a uniform distribution relative to the expected fidelity.
This provides a new \emph{expected distribution under noise} that the sampled distribution can be compared with.
Although this noise model is simplistic, it can be computed efficiently without substantial overhead and still provides reasonable results, as will be discussed later in \autoref{sec:evaluation}.

A final challenge for developers is determining the number of shots that should be used to sample the distribution.
A large number of shots requires drastic execution overhead, while a low number of shots may lead to inaccuracies when evaluating the assertion.
To address this challenge, the proposed framework also computes a recommended number of samples that should be used to accurately sample the program state.
This is achieved by simulating the \chisquare test with randomly sampled distributions to determine a number of samples that allows the test to consistently achieve acceptable $p$-values.

\section{Evaluation}
\label{sec:evaluation}

To evaluate the proposed framework, which is made available as an open-source implementation at~\redacted{\mbox{\url{https://github.com/munich-quantum-toolkit/debugger}}}, it was tested with a variety of quantum programs.
All evaluations were performed on the \texttt{ibm\_brisbane} device provided by the IBM Quantum Cloud~\cite{IBMQuantum}.
This device allows up to 127 qubits.
The results of these tests are summarized in this section.

\subsection{Assertion Evaluation for Small Programs}

To test the effectiveness of assertion evaluation on real quantum computers, a test set of 90 quantum programs was generated.
Each program implements a \emph{hardware-efficient ansatz}~\cite{kandala2017} (commonly used in variational quantum eigensolvers~\cite{peruzzo2014}) with 12, 14, or 16 qubits.
However, for each program, one error was introduced by removing a randomly selected instruction.
Each evaluated program ends with an equality assertion, determining whether the ansatz yields the expected outcome.
The equality assertion specifies its expected state through an explicit state vector representation.
After translating the assertions into measurements, the programs were executed on the \texttt{ibm\_brisbane} quantum computer over 8192 shots.
The measurement outcomes were then used to evaluate the assertion included in each of the programs.

To best evaluate the effectiveness of the assertion evaluation scheme, the programs were simultaneously executed using the decision-diagram-based simulator from the \mbox{Munich Quantum Toolkit's (MQT; \cite{mqt})} MQT Core library~\cite{burgholzer2025MQTCore}. %
As simulation allows perfect insight into the underlying state, the assertions did not  need to be translated into measurements for this purpose and were instead evaluated by directly accessing the simulation's state vector.
Due to this unrestricted access, the simulation results confirmed that all assertions are indeed able to detect the error.
Ideally, the same results should be obtained from the evaluations on the  \texttt{ibm\_brisbane} quantum computer, even in the presence of noise.

\begin{figure}
    \centering
    \includegraphics[width=0.48\textwidth]{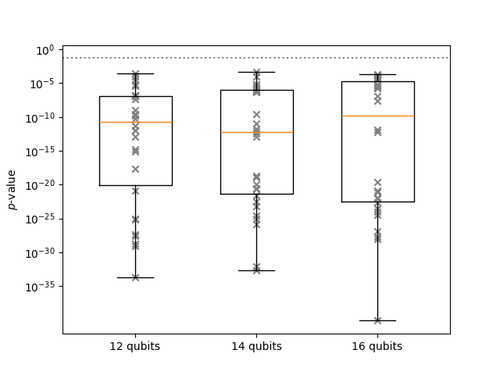}
    \caption{The $p$-values reported after evaluating a 90 incorrect hardware-efficient ansatz implementations with different numbers of qubits. 
    The dotted line marks a threshold of $0.05$ commonly employed when evaluating $p$-values.
    $p$-values below this threshold indicate that the assertion has detected an error.}
    \label{fig:eval-p}
    \vspace{-0.5cm}
\end{figure}

\autoref{fig:eval-p} illustrates the results of this evaluation.
All executed programs are grouped by the number of employed qubits.
The $p$-values reported by the verification procedure are then visualized using box plots, where each individual test result is marked as a cross.
Furthermore, the significance level $\alpha = 0.05$ is marked as a dotted line.
An assertion is considered satisfied, if its corresponding mark in the figure is located above the dotted line.
Therefore, as all marks are located below it, all evaluated assertions have successfully detected an error, even on a noisy quantum computer. 
The reported median values, all located around $10^{-10}$ are clearly below the required confidence level and, therefore, decisively indicate an error.
However, even the upper extremes of the reported $p$-values are well below the required confidence level, reaching a maximum of $4.6 \times 10^{-4}$.

The same test was also executed using the unmodified correct programs.
Each of these tests yielded a $p$-value between $0.9$ and $0.99$, indicating that the assertions did not falsely detect an error in correct programs.
\subsection{Large-Scale Assertion Evaluation}

While small quantum programs can often be tested efficiently, even on simulated quantum computers, increasing the program size quickly leads to a road block for simulation.
To test whether assertions evaluated on real quantum computers lead to an advantage in the search for errors, a larger quantum program containing an equality assertion was executed on the \texttt{ibm\_brisbane} device.
Due to the current limitations of quantum computers, such a program requires a low depth to prevent noise from fully obfuscating the program outcome.
Therefore, a simple program was constructed for this evaluation.
The program uses 32 qubits to create 16 Bell pairs.
As the individual Bell pairs do not interact with each other, a large number of instructions can be executed in parallel, keeping the depth minimal.
An assertion is then used to verify that two of these Bell pairs are created correctly.
An excerpt of the corresponding quantum program is illustrated in \autoref{lst:eval-large}\footnote{Clearly, such a program can be simulated efficiently even when using more than 32 qubits. While this evaluation aims to show the applicability of the proposed framework with a large number of qubits, it does not necessarily demonstrate an advantage over classical simulation.}.

\begin{lstfloat}[t]
    \begin{lstlisting}[language=mqt-qasm,caption={An excerpt of a program that creates 16 Bell pairs and tests the Bell pairs of \texttt{q[0]} and \texttt{q[1]} as well as \texttt{q[30]} and \texttt{q[31]}.\vspace{-0.2cm}},label=lst:eval-large, escapeinside={(*}{*)}]
qreg q[32];

h q[0];
cx q[0], q[1];

h q[2];
cx q[2], q[3];
...

h q[30];
cx q[30], q[31];

assert-eq q[0], q[1], q[30], q[31] = { 0.5, 0, 0, 0.5, 0, 0, 0, 0, 0, 0, 0, 0, 0.5, 0, 0, 0.5 }
    \end{lstlisting}
    \vspace{-0.3cm}
\end{lstfloat}

\begin{figure}
    \centering
    \includegraphics[width=\columnwidth]{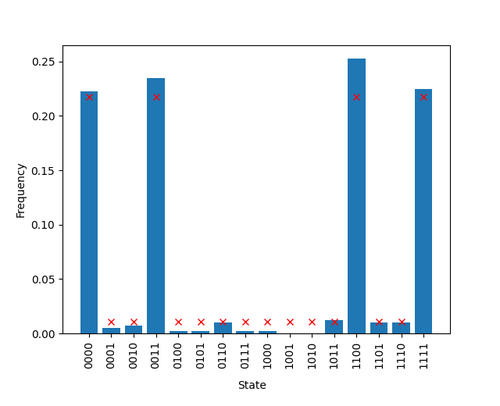}
    \caption{The frequencies of all measured states when running the program from \autoref{lst:eval-large} on a real quantum computer. 
    The crosses indicate the expected frequencies under an iid noise model.}
    \label{fig:eval-large}
    \vspace{-0.6cm}
\end{figure}

For the reliable evaluation of the assertion on Line~14, the proposed framework suggested the execution of 400 shots.
The measurement outcomes obtained after executing this program are illustrated in \autoref{fig:eval-large}. For each measured state,
it shows the frequency with which the state was recorded among the 400 shots as a bar, while the crosses indicate the expected frequency under the assumption of independent and identically distributed (iid) noise.

Running the proposed runtime verification method on the obtained outcomes resulted in a $p$-value of $0.9934$, indicating a high likelihood that the assertion is satisfied.
Performing the same evaluation using $1000$ shots yielded a $p$-value of $0.9996$, while an evaluation using $200$ shots instead resulted in a $p$-value of $0.7833$.
While, in this case, the latter approach still selects the correct assertion outcome, the proposed framework uses conservative performance estimates to reduce the likelihood of false outcomes.

To show that the proposed framework can also detect errors in larger programs, we modified the program by replacing the Hadamard gate on Line~11 with an \texttt{X} gate.
This prevents the \texttt{cx} gate in the subsequent line from entangling the qubits \texttt{q[30]} and \texttt{q[31]}.
Therefore, we expect a $p$-value below the significance level $\alpha = 0.05$ when evaluating this assertion.
Executing the modified program and evaluating the assertion yielded a $p$-value of $2.02 \times 10^{-6}$, indicating that the assertion is not satisfied.

This evaluation has shown that the proposed framework can also be applied to larger programs. 
Errors have been detected successfully, while correct behaviour did not lead to false positives.
However, it should be noted that the quality of the employed hardware is still a limiting factor for the performance of the proposed framework.
While low-depth programs as the one investigated in this evaluation can be executed with a low error rate, more complex programs may lead to levels of noise that cannot be mitigated by the techniques employed by the proposed framework.
Therefore, while this evaluation has shown promising results for low-depth problem instances, improvements in the quality of NISQ devices are still required for this framework to be universally applicable.

\vfill
\section{Conclusion}
\label{sec:conclusion}

In this work, we proposed a practical framework for evaluating assertions on real quantum computers.
This solution addresses three key challenges in the evaluation of quantum programs:
The automatic translation of abstract assertions into executable measurements,
the optimization of execution overhead caused by the introduced measurements,
and robust statistical verification using a \mbox{\chisquare test} to determine whether an assertion is satisfied, even in the presence of noise.
An open-source implementation of the framework is available at \redacted{\url{https://github.com/munich-quantum-toolkit/debugger}}.
Comprehensive evaluations demonstrate that the framework achieves results comparable to noiseless simulation in realistic noise environments. 
Moreover, even as program scale increased, the evaluations reliably detected assertion violations with minimal performance degradation.
This way, the proposed framework establishes a critical foundation for testing and debugging quantum programs at scales beyond classical simulation, enabling practical quantum software development in the near future.

\redacted[\nothing]{
  \vspace{-0.2cm}
\section*{Acknowledgments}
This work received funding from the European Research Council (ERC) under the
European Union's Horizon 2020 research and innovation program (grant agreement 
No. 101001318), was part of the Munich Quantum Valley, which is supported by 
the Bavarian state government with funds from the Hightech Agenda Bayern Plus, 
and has been supported by the BMWK on the basis of a decision by the German 
Bundestag through project QuaST, as well as by the BMK, BMDW, and the State of 
Upper Austria in the frame of the COMET program (managed by the FFG).
We acknowledge the use of IBM Quantum services for this work. 
The views expressed are those of the authors, and do not reflect the official policy or position of IBM or the IBM Quantum team.
}

\clearpage
\printbibliography

\end{document}